\documentclass[prb, 11pt,letterpaper,superscriptaddress,floatfix,nofootinbib,notitlepage]{revtex4-1}

\usepackage{textcomp}
\usepackage{amsmath}
\usepackage{amsfonts}
\usepackage{amssymb}
\usepackage{graphicx}
\usepackage{siunitx}
\usepackage{caption}
\usepackage{setspace}
\usepackage{xcolor}
\setcitestyle{super}
\usepackage[colorlinks, citecolor={blue}, linkcolor={black}]{hyperref}

\begin{document}
\raggedbottom

\author{Yuan Cao}
\thanks{These authors contributed equally}
\affiliation{Department of Physics, Massachusetts Institute of Technology, Cambridge, Massachusetts 02139, USA}

\author{$^{\!\!\!\!,\ \!\dagger}$ Jeong Min Park}
\thanks{These authors contributed equally}
\affiliation{Department of Physics, Massachusetts Institute of Technology, Cambridge, Massachusetts 02139, USA}

\author{$^{\!\!\!\!,\ \!\dagger}$ Kenji Watanabe}
\author{Takashi Taniguchi}
\affiliation{National Institute for Materials Science, Namiki 1-1, Tsukuba, Ibaraki 305-0044, Japan}

\author{Pablo Jarillo-Herrero}
\email{caoyuan@mit.edu}\email{parkjane@mit.edu}\email{pjarillo@mit.edu}
\affiliation{Department of Physics, Massachusetts Institute of Technology, Cambridge, Massachusetts 02139, USA}

\title{Large Pauli Limit Violation and Reentrant Superconductivity in Magic-Angle Twisted Trilayer Graphene}

\maketitle

\textbf{Moiré quantum matter has emerged as a novel materials platform where correlated and topological phases can be explored with unprecedented control. Among them, magic-angle systems constructed from two or three layers of graphene have shown robust superconducting phases with unconventional characteristics \cite{cao_unconventional_2018,yankowitz_tuning_2019,lu_superconductors_2019,park_tunable_2021,hao_electric_2021}. However, direct evidence for unconventional pairing remains to be experimentally demonstrated. Here, we show that magic-angle twisted trilayer graphene (MATTG) exhibits superconductivity up to in-plane magnetic fields in excess of \SI{10}{\tesla}, which represents a large ($2\sim3$ times) violation of the Pauli limit for conventional spin-singlet superconductors. This observation is surprising for a system which is not expected to have strong spin-orbit coupling. Furthermore, the Pauli limit violation is observed over the entire superconducting phase, indicating that it is not related to a possible pseudogap phase with large superconducting amplitude pairing. More strikingly, we observe reentrant superconductivity at large magnetic fields, which is present in a narrower range of carrier density and displacement field. These findings suggest that the superconductivity in MATTG is likely driven by a mechanism resulting in non-spin-singlet Cooper pairs, where the external magnetic field can cause transitions between phases with potentially different order parameters. Our results showcase the richness of moiré superconductivity and may pave a new route towards designing next-generation exotic quantum matter.}

A recent advance in quantum materials is the capability of creating artificial moiré superlattices via stacking two-dimensional materials with a twist angle and/or a lattice mismatch. In certain moiré superlattices, the appearance of nearly flat bands gives rise to a variety of correlated electronic phenomena\cite{cao_correlated_2018, cao_unconventional_2018, yankowitz_tuning_2019, lu_superconductors_2019, sharpe_emergent_2019, serlin_intrinsic_2020, chen_evidence_2019, burg_correlated_2019, shen_correlated_2020, cao_tunable_2020, liu_tunable_2020, regan_mott_2020, tang_simulation_2020, wang_correlated_2020, park_tunable_2021, hao_electric_2021}, including correlated insulators, ferromagnetic phases, and in particular, superconductivity. Robust superconductivity has been reproducibly found in magic-angle twisted bilayer graphene (MATBG)\cite{cao_unconventional_2018, yankowitz_tuning_2019,lu_superconductors_2019}, and more recently in magic-angle twisted trilayer graphene (MATTG)\cite{park_tunable_2021, hao_electric_2021}. The simultaneous presence of correlated insulator/resistive states in these systems has elicited extensive interest in the origin of this unusual superconducting phase. Moreover, MATTG exhibits a unique electric displacement field tunability and the superconducting state in MATTG can be tuned into the ultra-strong coupling regime. These aspects make MATTG an attractive platform to investigate the nature of moiré superconductivity.

One of the most fundamental questions about superconductivity is its pairing symmetry, namely the spatial symmetry and the spin configuration. The spatial symmetry can be classified as $s$-wave, $p$-wave, $d$-wave, or other exotic symmetries, whereas the spin configuration can be spin-singlet or spin-triplet. Most superconductors have a spin singlet pairing, including conventional superconductors described by Bardeen–Cooper–Schrieffer (BCS) theory, as well as even-parity unconventional superconductors, such as cuprates\cite{lee_doping_2006}. On the other hand, evidence of spin-triplet superconductivity has been found in only a few systems, such as UPt\textsubscript{3}\cite{strand_transition_2010,schemm_observation_2014} and UTe\textsubscript{2}\cite{ran_nearly_2019}. Spin-triplet pairing between fermionic atoms has long been investigated in the superfluid He-3, which exhibits a rich phase diagram consisting of different types of triplet phases\cite{leggett_theoretical_1975}. Beyond fundamental physics, additional interest in spin-triplet superconductivity has emerged because of the accompanying odd-parity spatial symmetry, which in some cases can host topological states\cite{kitaev_unpaired_2001} important for fault-tolerant quantum computing.

\begin{figure}
\includegraphics[width=1\textwidth]{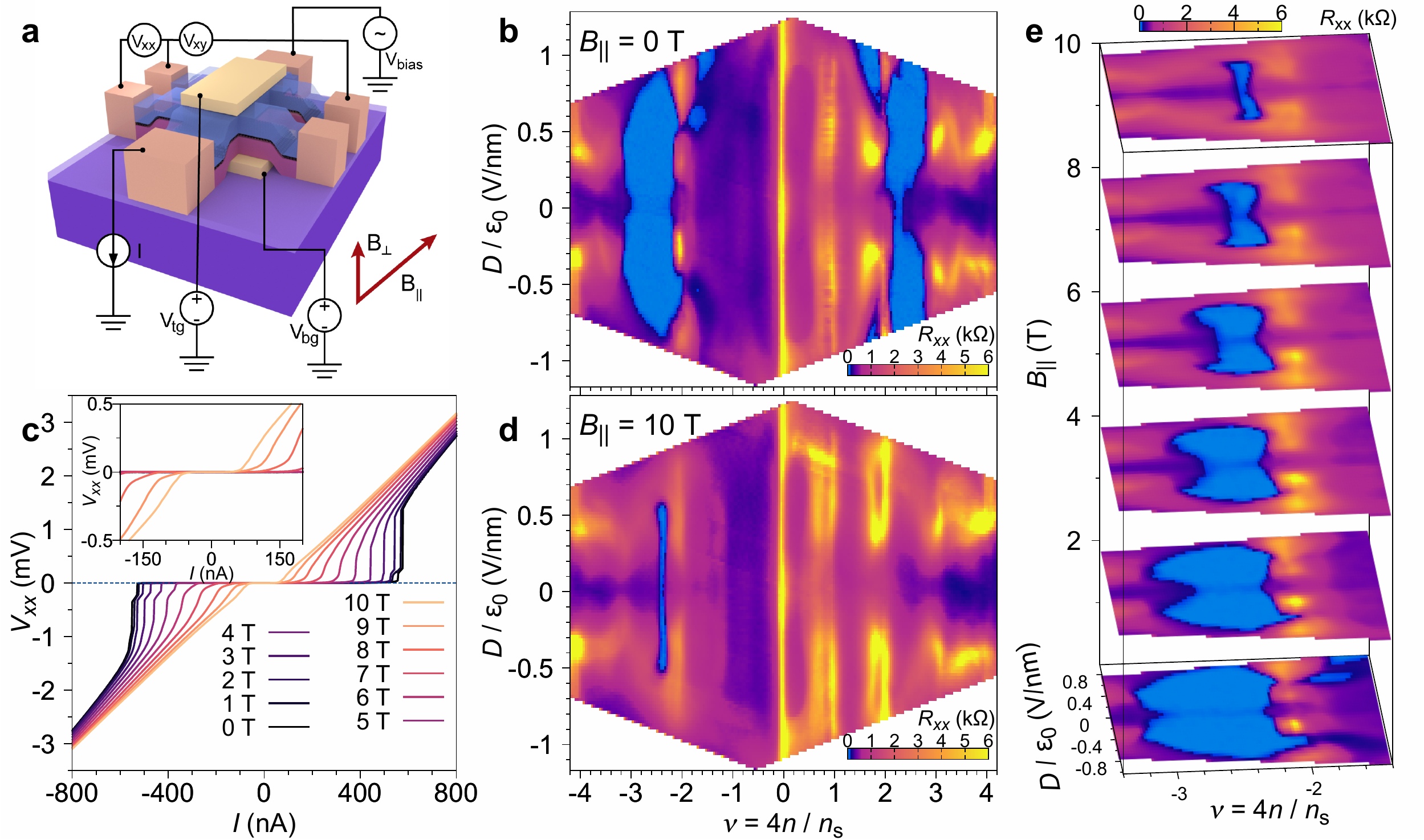}
\caption{Superconductivity in MATTG at high in-plane magnetic fields. (a) Experimental schematics. Four-probe measurements are performed by flowing current $I$ and measuring longitudinal voltage difference $V_{xx}$ and Hall voltage $V_{xy}$ as shown. Top and bottom gate voltages, $V_{tg}$ and $V_{bg}$, are applied to control the carrier density and electric displacement field in the sample. In-plane field $B_{\parallel}$ parallel to the 2-dimensional plane and and out-of-plane field $B_{\perp}$ perpendicular to the plane are shown. Small $B_{\perp}$ is applied to correct for a possible tilt of the sample with respect to $B_{\parallel}$. (b,d) Longitudinal resistance $R_{xx}$ at $B_{\parallel}=\SI{0}{\tesla}$ (b) and $B_{\parallel}=\SI{10}{\tesla}$ (d) at $T=\SI{300}{\milli\kelvin}$. (c) Voltage-current ($V$-$I$) curves versus $B_\parallel$ at $\nu=-2.4$, $|D/\varepsilon_0| = -\SI{0.44}{V/nm}$, and $T=\SI{300}{\milli\kelvin}$. Inset: zoom-in into the middle region shows the flatness of $V$-$I$ curves at large $B_{\parallel}$. (e) Evolution of superconductivity on the hole-doped side as a function of $B_{\parallel}$. Superconductivity persists up to $B_{\parallel}=\SI{10}{\tesla}$.} 
\end{figure}

In this Article, we perform quantum electronic transport measurements in MATTG in the presence of a magnetic field parallel to the sample plane to gain insight into the spin configuration of the superconducting state. Our results indicate that MATTG is unlikely to be a spin-singlet superconductor. We fabricated high-quality MATTG devices, where the adjacent layers are sequentially twisted by $\theta$ and $-\theta$, where $\theta\sim\SI{1.57}{\degree}$ is the magic-angle for MATTG, such that the topmost and bottommost layers are aligned, as detailed previously\cite{park_tunable_2021} (Fig. 1a). While theory predicts the formation of nearly flat bands in alternating twist angle trilayer graphene, similar to those in MATBG,  but at a new magic angle\cite{khalaf_magic_2019},  MATTG additionally exhibits a unique electric field tunability\cite{park_tunable_2021}. This structure can be reduced to MATBG-like flat bands and a dispersive Dirac band\cite{khalaf_magic_2019, carr_ultraheavy_2020, lei_mirror_2020, calugaru_tstg_2021}, whose hybridization can be further controlled by the application of electric displacement field $D$. The number of electrons per moiré unit cell, or the moiré filling factor, is defined by $\nu=4n/n_s$, where $n$ is the carrier density and $n_s=8\theta^2/\sqrt{3}a^2$ is the superlattice density ($a=\SI{0.246}{\nano\meter}$ is the lattice constant of graphene). Figure 1b shows the longitudinal resistance $R_{xx}$, as a function of $\nu$ and $D$ without any magnetic field, where a number of $D$-dependent correlated resistive states are present at integer filling factors $\nu=+1, \pm2, +3, \pm4$. Superconductivity appears in the vicinity of the $\nu=\pm2$ correlated states, and the highest critical temperature $T_{c}$ is found at $\nu=-2-\delta$ ($\delta$ is a small fraction smaller than one), approaching \SI{2.9}{\kelvin}\cite{park_tunable_2021,hao_electric_2021}. The critical temperature is further tunable by $D$, and the optimal $D/\varepsilon_0$ that maximizes $T_{c}$ is located in the range $\pm 0.4\sim$ \SI{0.5}{\volt\per\nano\meter}. Near this optimal point in doping and $D$, the superconductivity was found to be in the ultrastrong coupling regime\cite{park_tunable_2021}.

\section{Pauli limit violation}

In a generic superconductor, the application of an external magnetic field suppresses superconductivity mainly in two ways\cite{tinkham_introduction_2004}. One is via the formation of vortices, in a type-II bulk superconductor, which leads to loss of superconducting coherence when the average spacing between vortices is below their characteristic length $\xi$. However, such suppression is nearly absent when the magnetic field is parallel to the plane of an atomically thin two-dimensional superconductor. For example, the in-plane field required to thread one flux quantum laterally through the MATTG unit cell is well above \SI{100}{\tesla}. Alternatively, a magnetic field can suppress superconductivity via the Zeeman effect, or through in-plane orbital effects\cite{kwan_twisted_2020,cao_nematicity_2020}. The Zeeman effect, in particular, imposes an upper bound on the critical magnetic field of spin-singlet superconductors, known as the Pauli (or Clogston-Chandrasekhar) limit\cite{chandrasekhar_note_1962, clogston_upper_1962}, corresponding to $B_P=(\SI{1.86}{\tesla\per\kelvin})\times T_c$ for weakly-coupled BCS superconductors. Above this limit, the formation of Cooper pairs is energetically unfavorable. However superconductivity above the Pauli limit can still exist in the presence of finite-momentum pairing or strong spin-orbit coupling. The former gives rise to the Fulde-Ferrell-Larkin-Ovchinnikov (FFLO) state\cite{fulde_superconductivity_1964, larkin_nonuniform_1965}, which can boost the low temperature critical magnetic field beyond the Pauli limit by a small amount\cite{burkhardt_fulde-ferrell-larkin-ovchinnikov_1994}. The latter can lead, for example, to an Ising-like type of pairing, which can boost the critical field well beyond the Pauli limit\cite{lu_evidence_2015,saito_superconductivity_2016, xi_ising_2016}. For MATTG, the nominal Pauli limit at the optimal doping and displacement field is on the order of $4\sim $\SI{5}{\tesla}, depending on the selected resistance threshold.

\begin{figure}
\includegraphics[width=\textwidth]{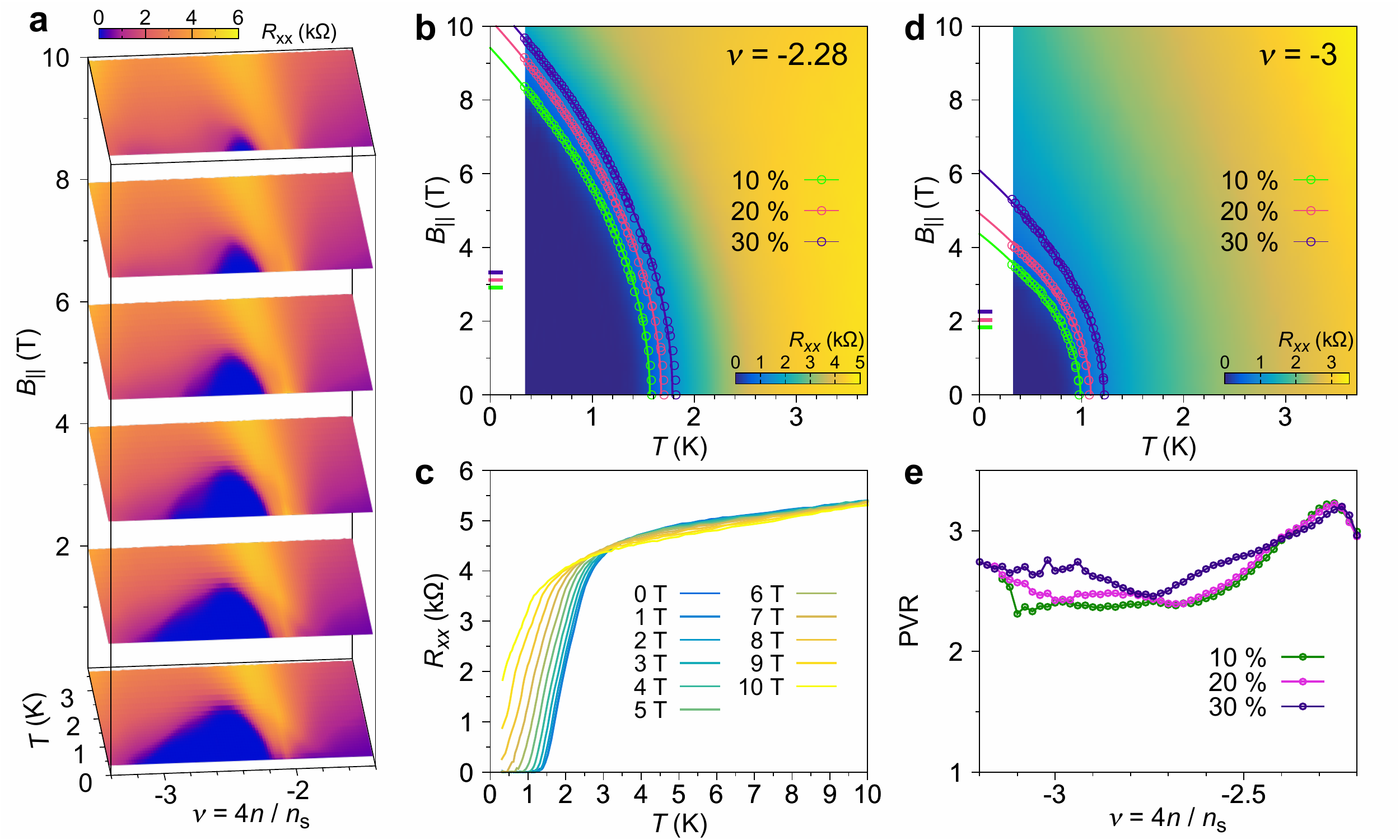}
\caption{Large Pauli limit violation in MATTG. (a) Evolution of the superconducting dome as a function of $B_\parallel$. Each color plane shows the resistance versus $\nu$ and $T$ at a fixed $B_\parallel$. At $\nu=-2.4$ (optimal doping), $T_c$ decreases to \SI{1.35}{\kelvin} when $B_\parallel=\SI{10}{\tesla}$. (b) $B_\parallel$-$T$ phase diagram at $\nu=-2.28$. The data points denote constant-resistance contours at \SI{10}{\percent}, \SI{20}{\percent}, and \SI{30}{\percent} of the zero-field normal state resistance, and the colored tickmarks on the $B_{\parallel}$-axis represent the corresponding Pauli limit. The contours roughly follow the Ginzburg-Landau expression $T\propto 1-\alpha B_\parallel^2$ (solid curves), where $\alpha$ is a fitting parameter. By extrapolating the contours to zero temperature, we find the critical magnetic fields \SI{9.41}{\tesla}, \SI{10.18}{\tesla}, \SI{10.73}{\tesla}, which give consistent Pauli limit violation ratios (PVR) of 3.23, 3.27, 3.23, for the \SI{10}{\percent}, \SI{20}{\percent}, and \SI{30}{\percent} contours, respectively. (c) Line cuts corresponding to (b) at a spacing of every \SI{1}{\tesla}. (d) Same plot as (b) for $\nu=-3$. The extraction of the PVR at this density yields 2.37, 2.42, and 2.69, using the three thresholds given above, respectively. (e) PVR extracted as a function of $\nu$ for the three resistance thresholds given above. All measurements above are taken at a displacement field $D/\varepsilon=\SI{-0.41}{V/nm}$.}  
\end{figure}

Strikingly, we find that the superconductivity in MATTG at $\nu=-2-\delta$ persists even in the presence of a large parallel magnetic field $B_\parallel=\SI{10}{\tesla}$ (Fig. 1d), much higher than the nominal Pauli limit. Figure 1e shows the evolution of the superconducting phase in the $\nu$-$D$ space as a function of $B_\parallel$. At $B_\parallel=\SI{10}{\tesla}$, a narrow strip near optimal doping, $\nu=-2.4$, remains superconducting, in the range $|D/\varepsilon_0| \leq \SI{0.6}{V/nm}$. In Fig. 1c, the robustness of superconductivity (i.e. not a normal state with low resistance) is shown by measuring the evolution of the $V_{xx}$-$I$ curves as a function of $B_\parallel$. While the critical current steadily decreases when $B_\parallel$ increases, it is clear that at $B_\parallel=\SI{10}{\tesla}$, $V_{xx}$-$I$ still exhibits an extremely flat region at finite dc current bias, indicating zero resistance, and a sharp peak in the differential resistance $dV_{xx}/dI$ at the critical current. 

In order to obtain the extent of Pauli limit violation, we investigate the multi-dimensional data sets in $\nu$, $B_\parallel$ and $T$ (displacement field is fixed at $D/\varepsilon_0=\SI{-0.41}{V/nm}$). As shown in Fig. 2a, the size of the $\nu$-$T$ superconducting dome at $-2-\delta$ shrinks as $B_\parallel$ is applied, where $T_{c}^{50\%}$ at $\nu=-2.4$ (the optimal doping) is reduced from \SI{2.7}{\kelvin} to \SI{1.35}{\kelvin} at $B_\parallel=\SI{10}{\tesla}$. The resistance versus $B_\parallel$-$T$ at $\nu=-2.28$ and the corresponding $R_{xx}$-$T$ line traces are shown in Fig. 2b and 2c, respectively. The constant-resistance contours in Fig. 2b correspond to roughly \SI{10}{\percent}, \SI{20}{\percent}, and \SI{30}{\percent} of the normal-state $R_{xx}$ at $B_\parallel=\SI{0}{\tesla}$, characterized by $R_{xx}$ at higher temperatures $T>\SI{4}{\kelvin}$, as shown in Fig. 2c. We find that the contours roughly follow the Ginzburg-Landau expression $T\propto 1-\alpha B_{\parallel}^2$ from $T_c$ down to the lowest temperature, $T\approx\SI{0.3}{\kelvin}$ (Fig. 2b), where $\alpha$ is a fitting parameter (see Methods). Using this formula, we can obtain the zero-temperature critical field $B_{c\parallel}(0)$ by extrapolation, using different percentages of the normal-state $R_{xx}$ to calculate $T_c$ (see Methods). At $\nu=-2.28$, we find $B_{c\parallel}(0)=$\SI{9.4}{T} (\SI{10}{\percent} normal resistance), whereas $T_c$ using the same threshold is \SI{1.56}{\kelvin} and the corresponding nominal Pauli limit is $B_P=\SI{2.9}{\tesla}$. The Pauli violation ratio (PVR), defined as $B_{c\parallel}(0)/B_P$, therefore reaches $3.2$ at this density. 

Figure 2e shows the PVR versus $\nu$ in the $-2-\delta$ dome extracted using the same method. Remarkably, the PVR is always above  $2$ over the entire range of $\nu$ in the $-2-\delta$ superconducting dome, and extractions with different normal-state $R_{xx}$ percentage thresholds give largely consistent values of PVR (see Extended Data Figure 1 for the electron-doped side, whose PVR values are also above 2). In particular, we find the PVR to be approximately $2.5$ even when $\nu$ is as large as $-3$, near the edge of the superconducting dome (see Fig. 2a, $B_\parallel=0$ slice and Fig. 2d), where $T_{c,10\%} < \SI{1}{\kelvin}$. Around this density and displacement field, the coupling strength determined by the coherence length and $T_c/T_F$ ($T_F$ is the Fermi temperature) is substantially smaller\cite{park_tunable_2021} than at $\nu=-2.28$. These observations indicate that the high PVR is not directly correlated with the superconducting coupling strength, but could rather be an inherent property of the superconducting state, such as its  spin configuration. 

\section{Reentrant Superconductivity}

\begin{figure}
\includegraphics[width=\textwidth]{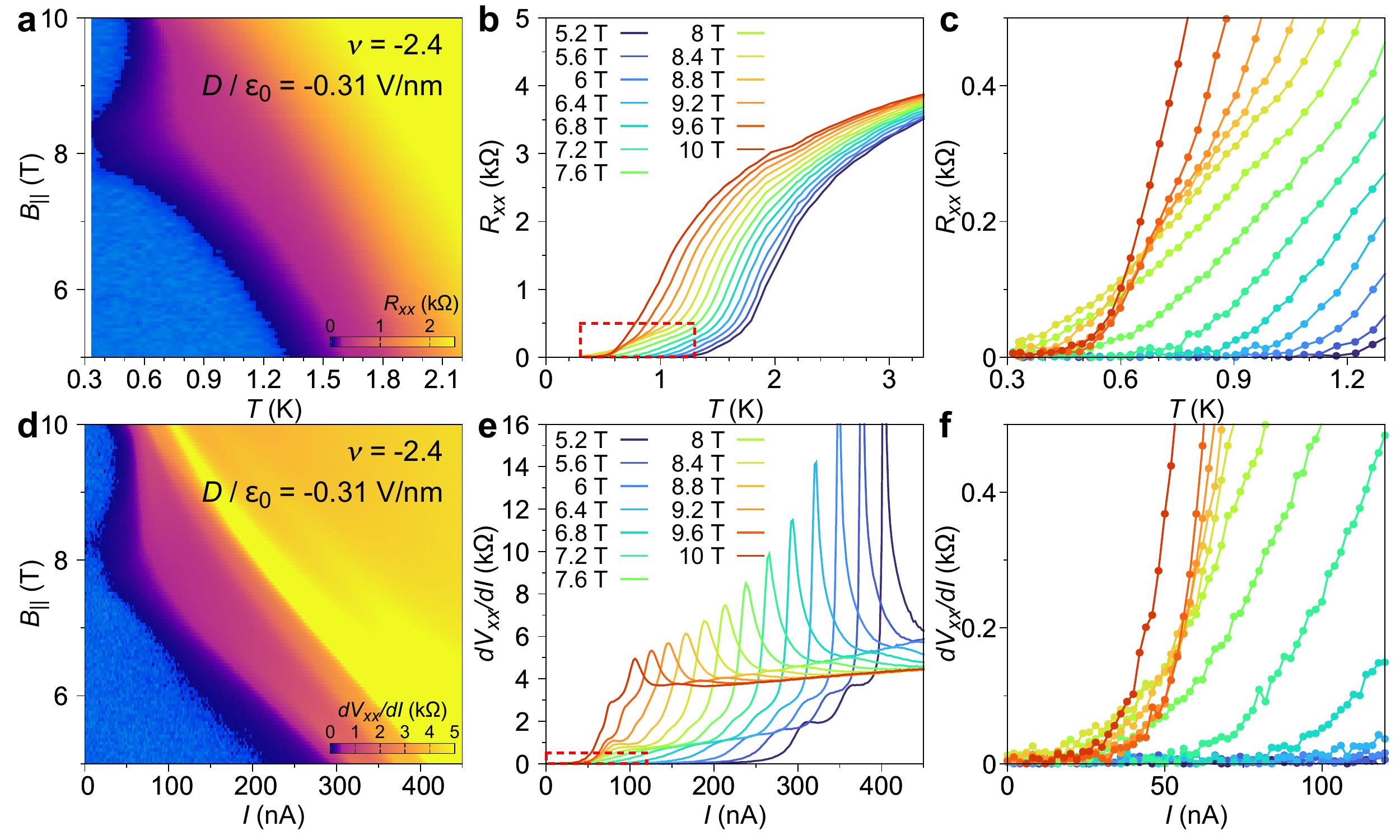}
\caption{Reentrant superconductivity. All data are taken at $\nu=-2.4$ and $D/\varepsilon_0=\SI{-0.31}{\volt\per\nano\meter}$. (a) $R_{xx}$ versus $B_{\parallel}$ and $T$, in the $B_{\parallel}$ range of $5\sim $\SI{10}{\tesla}. The superconducting phase present at $B_\parallel=0$ gets suppressed around $B_{\parallel}\sim\SI{8}{\tesla}$, where a reentrant superconducting phase starts appearing. The reentrant behaviour only exists in the near zero resistance region. (b-c) Linecuts of $R_{xx}$ versus $T$ (b) and zoom-in (c) as a function of $B_\parallel$ show the that the $R_{xx}-T$ curves show a non-monotonic behaviour near the BKT transition temperature, around the transition field of $B_{\parallel}\sim\SI{8}{\tesla}$. (d) Differential resistance $dV/dI$ versus $B_{\parallel}$ and $I$ shows a similar trend as the $T$-dependence, with the reentrant phase boundary near the critical current defining the zero resistance. (e-f) Linecuts of $dV/dI$ (e) versus $I$ and zoom-in (f) as a function of $B_\parallel$ show similar non-monotonic behaviour as in (b-c), demonstrating the reentrant superconducting phase. Color legends of (c),(f) are the same as (b) and (e), respectively.} 
\end{figure}

Beyond the large Pauli limit violation, we also observe additional superconducting phases at large $B_\parallel$, for a certain range of $D$ smaller than the optimal value. Figure 3a shows the high-field part ($B_\parallel>\SI{5}{\tesla}$) of the $B_\parallel$-$T$ phase diagram at $\nu=-2.4$ and $D/\varepsilon_0=\SI{-0.31}{\volt\per\nano\meter}$. In the low-temperature region, \emph{i.e.} close to the Berezinskii–Kosterlitz–Thouless (BKT) transition of the system\cite{park_tunable_2021}, the zero-resistance state (light blue color) disappears at around $B_\parallel=\SI{8}{\tesla}$. Above \SI{8}{\tesla}, however, a separate zero-resistance region reappears and persists to higher than \SI{10}{\tesla}, signaling a reentrant superconducting phase. On the other hand, $T_c^{50\%}$, which represents the center of the superconducting transition, decreases monotonically with the in-plane magnetic field. Figures 3b-c shows the corresponding $R_{xx}$-$T$ line cuts as $B_\parallel$ varies, where the reentrant superconductivity is manifested as the crossing of $R_{xx}$-$T$ curves around $T\approx\SI{0.6}{\kelvin}$. We observe a similar reentrant behaviour by examining the differential resistance versus dc current bias at high $B_\parallel$, as shown in Fig. 3d-f. For each curve in Fig. 3e and 3f, the differential resistance has a large peak corresponding to the major step in the $V_{xx}-I$ curve at the critical current (Fig. 1d), and multiple `shoulders' at smaller bias currents. We find the reentrant behaviour with respect to $B_\parallel$ at the first shoulder, which corresponds to a transition between a nondissipative state ($dV_{xx}/dI=0$, light blue region in Fig. 3d) and a slightly dissipative state ($dV_{xx}/dI>0$), whereas the position of the large $dV_{xx}/dI$ peak evolves monotonically with $B_\parallel$, analogous to the behaviour of $T_c^{50\%}$.

\begin{figure}
\includegraphics[width=\textwidth]{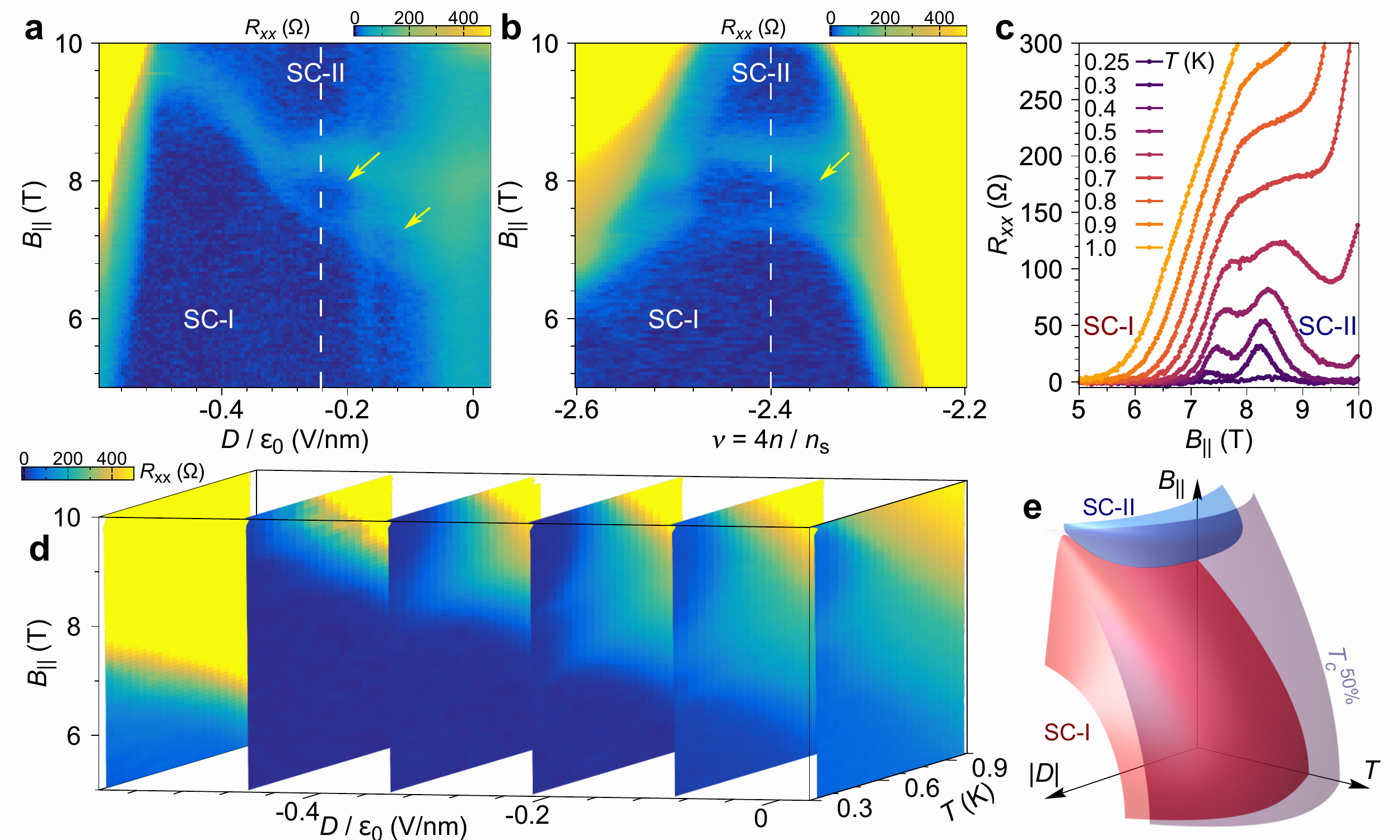}
\caption{Field-induced transition between superconducting phases in MATTG. (a) Resistance versus $D$ and $B_\parallel$ at optimal doping $\nu=-2.4$. Measurement is performed at $T=\SI{0.4}{\kelvin}$. SC-I and SC-II denote the zero-field superconducting phase (Fig. 2) and the high-field reentrant phase (Fig. 3), respectively. They are clearly separated by a resistive `filament'. At fine-tuned displacement fields of around \SI{-0.23}{\volt\per\nano\meter} and \SI{-0.12}{\volt\per\nano\meter}, we find additional regions with lower resistance (denoted by yellow arrows), which might signal the onset of additional phases. (b) Resistance versus $\nu$ and $B_\parallel$ at $D/\varepsilon_0=\SI{-0.24}{\volt\per\nano\meter}$ (dashed line in (a)). Here the white dashed line denotes the optimal doping $\nu=-2.4$. The yellow arrow denotes the additional superconducting region similar as in (a). (c) Temperature dependence of $R_{xx}$ versus $B_\parallel$, at $D$ and $\nu$ marked by the white dash line in (a) and (b). Between SC-I and SC-II, there are clearly two resistive peaks. The curves are measured when ramping down the field. Bidirectional sweep in $B_\parallel$ reveals some hysteretic behaviours, which might point towards a first-order transition (see Extended Data Figure 3). (d) Evolution of the $B_\parallel$-$T$ phase diagrams as $D$ is varied. The critical $B_\parallel$ where the transition between SC-I and SC-II occurs varies with $D$.  (e) Three-dimensional schematic of the phase diagram of the superconducting phases in the $|D|-B-T$ space. The red and blue surfaces denote the boundaries of the SC-I and SC-II phases, respectively, and the purple surface denotes the mean-field $T_c$ determined by \SI{50}{\percent} of the normal-state resistance.} 
\end{figure}

Investigating the reentrant superconducting behaviour in the full space of $\nu$, $D$, and $B_{\parallel}$ further reveals an intricate phase diagram with multiple superconducting phases. Figure 4a shows the resistance map versus $B_\parallel$-$D$. Here, we denote the prominent low-field and high-field zero-resistance regions as SC-I and SC-II. At large $D$ ($|D|/\varepsilon_0>\SI{0.5}{\volt\per\nano\meter}$), SC-I directly transitions to a dissipative state as $B_\parallel$ is increased. On the other hand, at intermediate $D$ ($\SI{0.3}{\volt\per\nano\meter}<|D|/\varepsilon_0<\SI{0.5}{\volt\per\nano\meter}$), we find that SC-I and SC-II are separated by a continuous resistive `filament-like' looking region, marking the boundary between these two phases. Interestingly, as $D$ decreases, smaller islands of superconducting regions appear between $7\sim $\SI{8}{\tesla}, as denoted by the yellow arrows. These islands may indicate the onset of additional reentrant states, which could perhaps be attributed to an admixture state of the SC-I and SC-II phases, or could also be a signature of new superconducting phases that only exist in a narrow range at finite $B_\parallel$. Figure 4b shows the density dependence of the various reentrant phases at $D/\varepsilon_0=\SI{-0.24}{\volt\per\nano\meter}$. All of the high-field superconducting phases are visible only in the range $-2.52<\nu<-2.28$, \emph{i.e.} close to the optimal doping $\nu\approx -2.4$. Figure 4c shows the temperature dependence of the transition between SC-I and SC-II. From \SI{0.3}{\kelvin} to \SI{0.6}{\kelvin}, a double-peak transition can be clearly observed. When the temperature is further lowered, the resistive features between the superconducting phases become weaker (see also Extended Data Figure 3a). We have performed bidirectional sweeps in $B_\parallel$ at this $\nu$ and $D$, which reveals a hysteretic behaviour and may point towards a first-order field-induced transition (see Extended Data Figure 3b,c). Figure 4d shows the evolution of the phases as a function of $T$ and $B_{\parallel}$ for different $D$ at fixed $\nu=-2.4$, which captures the evolution of $B_{\parallel}$ at which the reentrant transition between SC-I and SC-II occurs. The schematic phase diagram is summarized in Fig. 4e, where the phase boundaries of SC-I (red), SC-II (blue), and the transition defined by $T_{c,50\%}$ (purple) are illustrated. 

\section{Discussion}

Our results indicate that the spin configuration of the superconducting state in MATTG is unlikely to consist of spin singlets. Although large violations of the Pauli limit have been observed for spin-singlet superconductors, these are typically due to one of the following mechanisms: (i) Superconductors with strong spin-orbit coupling (SOC), such as the 2-dimensional transition metal dichalcogenides\cite{lu_evidence_2015,saito_superconductivity_2016, xi_ising_2016}. However, graphene and graphene multilayers are known for their very weak SOC, including both the intrinsic term as well as the Rashba term (for electric fields up to \SI{0.8}{\volt\per\nano\meter}, as we use here)\cite{avsar_colloquium_2020}, and to our knowledge there is no evidence to date that MATTG has a substantially enhanced SOC. Further theoretical and experimental work is necessary to establish if there is any appreciable contribution to the PVR from SOC in MATTG. (ii) Strongly-coupled superconductors (with large $T_{c}/T_{F}$ ratio) can exhibit a large $\Delta/k_{B}T_{c}$ ratio, where $\Delta$ is the magnitude of the pairing gap, which can give rise to a large apparent ratio of $B_c/T_c$ that exceeds the Pauli limit. In MATTG, however, the PVR does not exhibit significant variation across the entire superconducting dome (Fig. 2e), while the coupling strength varies by more than an order of magnitude\cite{park_tunable_2021}. It is therefore difficult for the strong-coupling mechanism to account for the Pauli limit violation across the entire superconducting dome. (iii) In a singlet superconductor, the FFLO state could be stabilized at high magnetic fields beyond the Pauli limit (at zero temperature). However, in a two-dimensional BCS superconductor, the $B_c$ enhancement is at most $\sim$\SI{40}{\percent} above the Pauli limit\cite{burkhardt_fulde-ferrell-larkin-ovchinnikov_1994}, much less than the three-fold violation observed in our experiments. Furthermore, the superconducting coherence that exists at fields beyond the Pauli limit is present even at finite temperature. As shown in Fig. 2b and 2d, the critical contours in the $B$-$T$ phase diagrams follow the quadratic behaviour from low temperature up to $T_c$. This implies that even close to $T_c$, where an FFLO state is unlikely to form, the critical $B$ is much higher than the expected value for a spin-singlet BCS superconductor. Therefore, while the FFLO-type physics could still be relevant to the high-field phases, it is difficult to fully account for the large Pauli limit violation in MATTG. 

Given that none of the usual mechanisms that lead to Pauli limit violation in spin-singlet superconductors are likely to play a dominant role in MATTG, it is natural to consider the possibility of a spin-triplet order parameter in MATTG. In a spin-triplet superconductor, the Cooper pairs have spin angular momentum $S=1$, and the spin configuration of the order parameter can be represented by a complex vector $\Vec{d}$ defined on the Fermi surface\cite{leggett_theoretical_1975,sigrist_introduction_2005}. The response of a spin-triplet state to an external $B$-field crucially depends on the alignment between $\Vec{d}$ and $B$. Neglecting orbital effects and SOC, an equal-spin pairing (ESP) state that has spins along the direction of $B$ ($\Vec{d}\cdot\Vec{B}=0$) does not respond to the field at all, whereas a non-ESP state with $\Vec{d}\parallel\Vec{B}$ is maximally suppressed by $B$, similar to a spin-singlet state. In magic-angle graphene systems, the additional valley degree of freedom can lead to an extra pair-breaking effect due to orbital effects, when Cooper pairs formed from electrons with opposite momenta (and thus opposite valleys) are subject to an in-plane $B$-field\cite{cao_nematicity_2020}. Regardless of the Cooper pairs spin configuration, this orbital pair-breaking effect can eventually lead to the suppression of superconductivity. Therefore, the ESP triplet state might be a viable candidate pairing state which could account for the large Pauli limit violation in the low-field state (SC-I), with an eventual possible suppression due to the orbital pair-breaking effect. An alternative scenario is a spin-valley locked pairing state at zero field\cite{khalaf_symmetry_2020}, which rotates into a spin-polarized state at high fields.

In addition to the large PVR, the observation of reentrant superconducting phases provides a further argument for a non-spin-singlet pairing. To date, superconductivity at high magnetic fields has most notably been identified in organic\cite{uji_magnetic-field-induced_2001, balicas_superconductivity_2001} and ferromagnetic\cite{aoki_review_2019, strand_transition_2010,schemm_observation_2014, ran_nearly_2019} superconductors. Different mechanisms have been suggested to explain these exotic phases, including dimensional
cross-over\cite{uji_magnetic-field-induced_2001}, exchange stabilization\cite{balicas_superconductivity_2001}, and ferromagnetic fluctuations\cite{aoki_review_2019,ran_nearly_2019}. Compared to these systems, we note that the low-field phase and the reentrant phase in MATTG are possibly separated by a first-order transition (see Methods and Extended Data Fig. 3), which is reminiscent of the transition between the A and B phases in He-3\cite{leggett_theoretical_1975}. We also note that the reentrant behaviour is only observed near the BKT transitions, and not in the higher temperature region where the initial drop in resistance occurs, suggesting that all the identified superconducting phases could correpsond to the same instability stemming from the normal phase \cite{leggett_theoretical_1975}. The change in $T_{BKT}$ near the transition might then be attributed to the difference in phase stiffness. As in the case of superfluid He-3\cite{leggett_theoretical_1975}, such behaviour could imply that both SC-I and SC-II are spin-triplet phases with different order parameters. As one possibility, while SC-I is the ground state at zero field, SC-II could be a spin-polarized phase (non-unitary) that is only stabilized at high magnetic fields. We point, however, that the presence of the additional valley degree of freedom in MATTG allows for a richer set of combinations of spin, valley, and spatial symmetries with multitude of possible order parameters, and further experimental and theoretical investigations are needed to determine the full pairing symmetry in the different superconducting phases of MATTG\cite{khalaf_symmetry_2020, christos_superconductivity_2020, cea_coulomb_2021}.

\section*{Acknowledgements}
We thank Liang Fu, Francisco Guinea, Steve Kivelson, Patrick Lee, Senthil Todadri, and Ashvin Vishwanath for fruitful discussions. 

This work has been primarily supported by the US Department of Energy (DOE), Office of Basic Energy Sciences (BES), Division of Materials Sciences and Engineering under Award DE-SC0001819 (J.M.P.). Help with transport measurements and data analysis were supported by the National Science Foundation
(DMR-1809802), and the STC Center for Integrated Quantum Materials (NSF Grant No. DMR-1231319) (Y.C.). P.J-H acknowledges support from the Gordon and Betty Moore Foundation's EPiQS Initiative through Grant GBMF9643. P.J-H acknowledges partial support by the Fundación Ramon Areces. The development of new nanofabrication and characterization techniques enabling this work has been supported by the US DOE Office of Science, BES, under award DE‐SC0019300. K.W. and T.T. acknowledge support from the Elemental Strategy Initiative conducted by the MEXT, Japan, Grant Number JPMXP0112101001, JSPS KAKENHI Grant Numbers JP20H00354 and the CREST(JPMJCR15F3), JST. This work made use of the Materials Research Science and Engineering Center Shared Experimental Facilities supported by the National Science Foundation (DMR-0819762) and of Harvard's Center for Nanoscale Systems, supported by the NSF (ECS-0335765).

\section*{Methods}

\subsection{Sample fabrication}

Briefly, the MATTG trilayer stack is sandwiched between two hBN flakes \SIrange{30}{80}{\nano\meter} thick. The hBN and graphene flakes were first exfoliated on SiO\textsubscript{2}/Si substrates and screened with optical microscopy. Then we use a dry pick-up technique to fabricate the multilayer stack. A layer of poly(bisphenol A carbonate)(PC)/polydimethylsiloxane (PDMS) on a glass slide is used to sequentially pick up the flakes. The three pieces of graphene that makes up MATTG are \emph{in situ} laser-cut from a single graphene flake\cite{park_flavour_2020}. The resulting stack is released on hBN on Pd/Au stack. The Hall-bar is defined with electron beam lithography and reactive ion etching. The top gate and electrical contacts are evaporated and lift off using Cr/Au. Refer to Ref. \cite{park_tunable_2021} for further details.

\subsection{Measurement and Data Analysis}

The electronic transport of MATTG is measured in a cryostat with base temperature of \SI{0.25}{\kelvin}. We bias the sample with an ac current with frequency \SIrange{10}{20}{\hertz}, and measure the four-probe voltage with SR-830 lock-in amplifiers, synchronized at the same frequency. The current and voltage signals are first amplified by \SI{1e7}{V/A} and \SI{1000}, respectively. Data of resistance hysteresis between SC-I and SC-II are measured using a current bias of \SI{5}{\nano\ampere}, while all other measurements are performed with \SI{1}{\nano\ampere} bias. For dc measurements (Fig. 1c, Fig. 3d-f), we use a digital-analog converter to provide the dc bias current. The dc voltage is measured using digital multimeter, while the differential resistance $dV_{xx}/dI$ is simultaneous recorded from the lock-in amplifier.

The in-plane field measurement is performed in a triple-axis vector magnet. We mounted the sample vertically so that an in-plane field up to \SI{10}{\tesla} could be applied using the Z axis of the magnet. The X axis magnet is used to compensate for the tilt of the sample (about 3 degree), so that the net field experienced by the sample is parallel to the sample with residual perpendicular field less than \SI{5}{\milli\tesla}.

For the graphics in Fig. 2b,d, Fig. 3a, and Extended Data Fig. 1 and 2, since the raw data were taken in a non-regular interval in temperature, we first interpolated the data into a regular grid in $B_\parallel$ and $T$ before plotting. We have checked that no artifact is introduced in this interpolation. 

\subsection{Pauli violation ratio (PVR) extraction}

For the PVR determination, we first extract the zero-field normal state resistance by fitting the high-temperature part of the data with a straight line $R_N(T)=aT+b$, where $a$ and $b$ are parameters. For a given threshold $p$ ($p=\SI{10}{\percent},\SI{20}{\percent},\SI{30}{\percent}$), we find the intersection of the zero-field resistance curve with $p\cdot R_N(T)$. The resistance at this intersection is denoted $R_N^p$. The intersection also defines the zero-field critical temperature $T_c^p(0)$. The data points in Fig. 2b and 2d are constant-resistance contours corresponding to $R_N^{\SI{10}{\percent}}$, $R_N^{\SI{20}{\percent}}$ and $R_N^{\SI{30}{\percent}}$ respectively.

Since each contour roughly follows $T= T_c^p(0)(1-\alpha_p B_\parallel^2)$ from $T_c^p(0)$ down to lowest temperature we can measure, we fit the points in each contour to this formula ($\alpha_p$ is a fitting parameter) and obtain the zero-temperature critical field through extrapolation, towards the point where the contour would intercept with the $T=0$ axis. This is given by $B_c^p(0)=\alpha_p^{-1/2}$. The corresponding PVR is then calculated as $\mathrm{PVR}=B_c^p(0)/[\SI{1.86}{\tesla\per\kelvin} T_c^p(0)]$. We performed this procedure independently for each $\nu$ and $p=10\%, 20\%, 30\%$ and plotted the result in Fig. 2e. 

\subsection{Additional Pauli limit violation data}

We observed large Pauli limit violation in other superconducting regions of the main device under study, as well as two other devices. Extended Data Fig. 1 shows the Pauli limit violation at representative densities and displacement fields in the $\nu=+2-\delta$ and $\nu=+2+\delta$ superconducting domes, on the electron-doping side of charge neutrality. From the \SI{10}{\percent}, \SI{20}{\percent}, and \SI{30}{\percent} normal-state resistance contours, we extract critical magnetic fields $B_c^{10\%}(0)=\SI{3.99}{\tesla}$, $B_c^{20\%}(0)=\SI{4.39}{\tesla}$ and $B_c^{30\%}(0)=\SI{4.93}{\tesla}$ for $\nu=+2-\delta$ (Extended Data Fig. 1a), and $B_c^{10\%}(0)=\SI{7.24}{\tesla}$, $B_c^{20\%}(0)=\SI{8.31}{\tesla}$ and $B_c^{30\%}(0)=\SI{10.45}{\tesla}$ for $\nu=+2+\delta$ (Extended Data Fig. 1b). These values correspond to PVR of 3.44, 2.98, 2.83 for $+2-\delta$ and 2.49, 2.37, 2.65 for $+2+\delta$, extracted using \SI{10}{\percent}, \SI{20}{\percent}, and \SI{30}{\percent}, respectively.

Extended Data Figure 2 shows the Pauli limit violation in two other MATTG devices we measured, device B and device C (the main device shown in the main text is denoted device A). Following the same extraction procedure as above, we obtain zero-temperature critical magnetic fields $B_c^{10\%}(0)=\SI{2.87}{\tesla}$, $B_c^{20\%}(0)=\SI{3.04}{\tesla}$ and $B_c^{30\%}(0)=\SI{3.25}{\tesla}$ for device B (Extended Data Fig. 2a), and $B_c^{10\%}(0)=\SI{3.35}{\tesla}$, $B_c^{20\%}(0)=\SI{3.46}{\tesla}$ and $B_c^{30\%}(0)=\SI{3.56}{\tesla}$ for device C (Extended Data Fig. 2b). This gives PVR of 2.13, 2.00, and 2.00 for device B, and 2.29, 2.23, and 2.19 for device C, extracted using \SI{10}{\percent}, \SI{20}{\percent}, and \SI{30}{\percent}, respectively.

Combining these data, we conclude that the large Pauli limit violation (two-fold to three-fold) is a universal feature in MATTG superconductivity. This suggest that the Pauli limit violation is likely an intrinsic property of the superconductivity, and may point towards an unconventional spin configuration, as discussed in the main text.

\subsection{Field-induced transition between SC-I and SC-II}

To further probe the nature of the transition between SC-I and SC-II phases, we performed bidirectional sweeps in $B_\parallel$ while fixing $\nu$, $D$, and $T$. We find that the resistance measured while scanning up $B_\parallel$ is considerably different than the resistance measured while scanning down, showing a hysteresis. However, the behaviour appears to be very sensitive to the measurement environment and varies from scan to scan. Extended Data Figure 3b,c shows two such scans at $\nu=-2.4$, $D/\varepsilon_0=\SI{-0.24}{\volt\per\nano\meter}$, and $T=\SI{0.3}{\kelvin}$. The only difference between the two scans is that the BNC cables that connect from the cryostat to the lock-in amplifiers are rearranged. In the first scan, the scan-up and scan-down curves are clearly offset in $B_\parallel$. In the second scan, the peak amplitude shows a considerable difference, while no offset in $B_\parallel$ is seen. 

We believe that the observation of hysteretic behaviour as well as the extreme sensitivity to environmental disturbances (rearrangement of cables could alter the coupling to external electromagnetic noise in the laboratory) is a strong evidence that the transition is of a first-order nature. For example, one possible scenario is that SC-I and SC-II are of different spin-triplet order parameters, and SC-II is stabilized by a high magnetic field and separated from SC-I by a first-order transition, similar to the A and B phases in He-3. Further studies are necessary to determine the precise nature of these phases.

\setcounter{figure}{0}
\renewcommand{\figurename}{Extended Data Fig.}

\clearpage

\section{Extended Data Figures}

\begin{figure}[!ht]
\includegraphics[width=\textwidth]{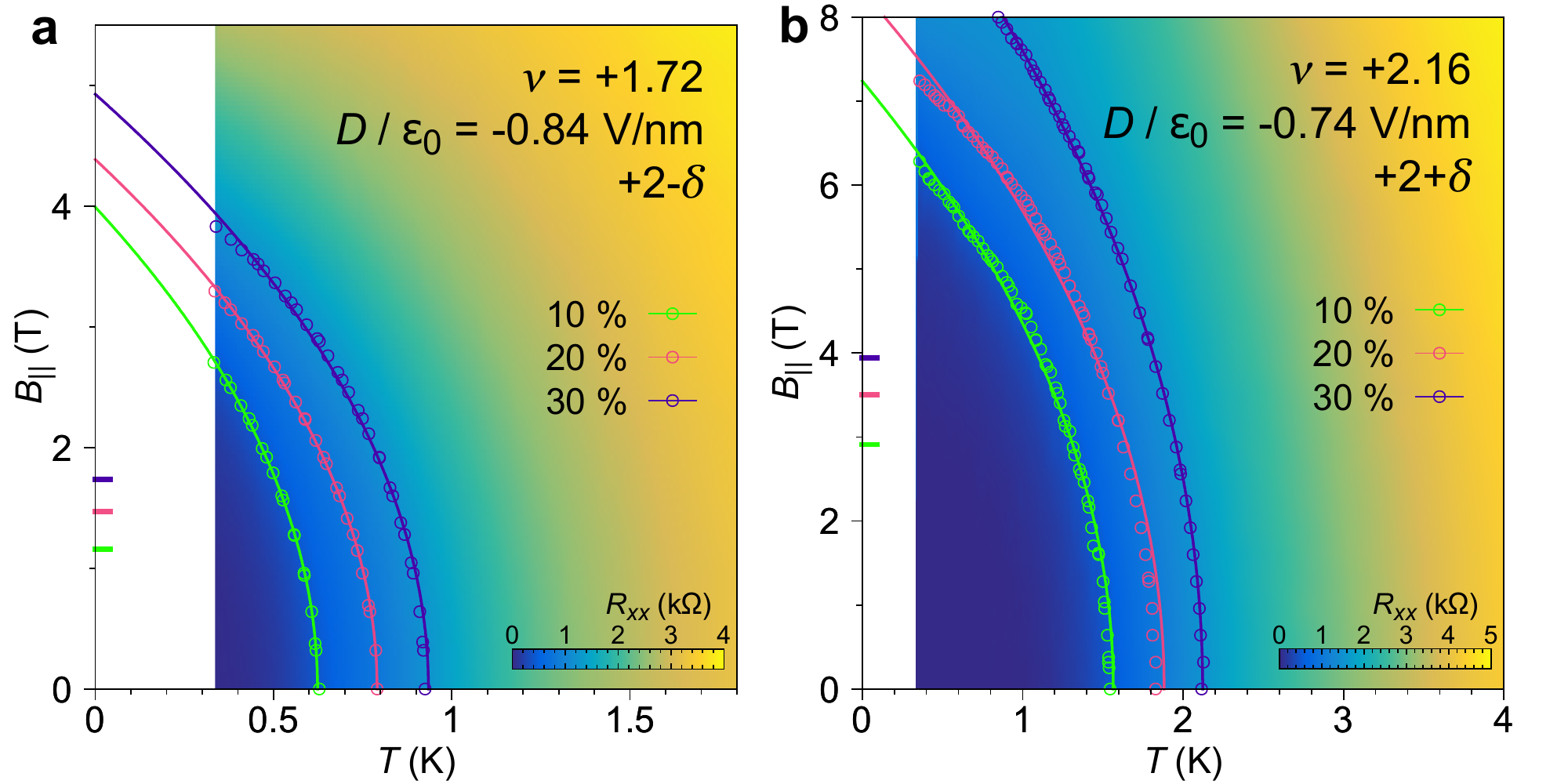}
\caption{Pauli limit violation for electron doping. (a) $B_\parallel$-$T$ phase diagram at a density in the $+2-\delta$ superconducting dome. The extracted Pauli limit violation ratios using \SI{10}{\percent}, \SI{20}{\percent} and \SI{30}{\percent} of normal resistance as the threshold are 3.44, 2.98, and 2.83 respectively. (b) $B_\parallel$-$T$ phase diagram at a density in the $+2+\delta$ superconducting dome. The extracted Pauli limit violation ratios using \SI{10}{\percent}, \SI{20}{\percent} and \SI{30}{\percent} of normal resistance as the threshold are 2.49, 2.37, and 2.65 respectively. The solid lines show the fit to the Ginzburg-Landau expression $T\propto 1-\alpha B_\parallel^2$, and the color tickmarks at $T=0$ show the corresponding Pauli limit, the same as in Fig. 2 in the main text.}
\end{figure}

\clearpage
\begin{figure}[!ht]
\includegraphics[width=\textwidth]{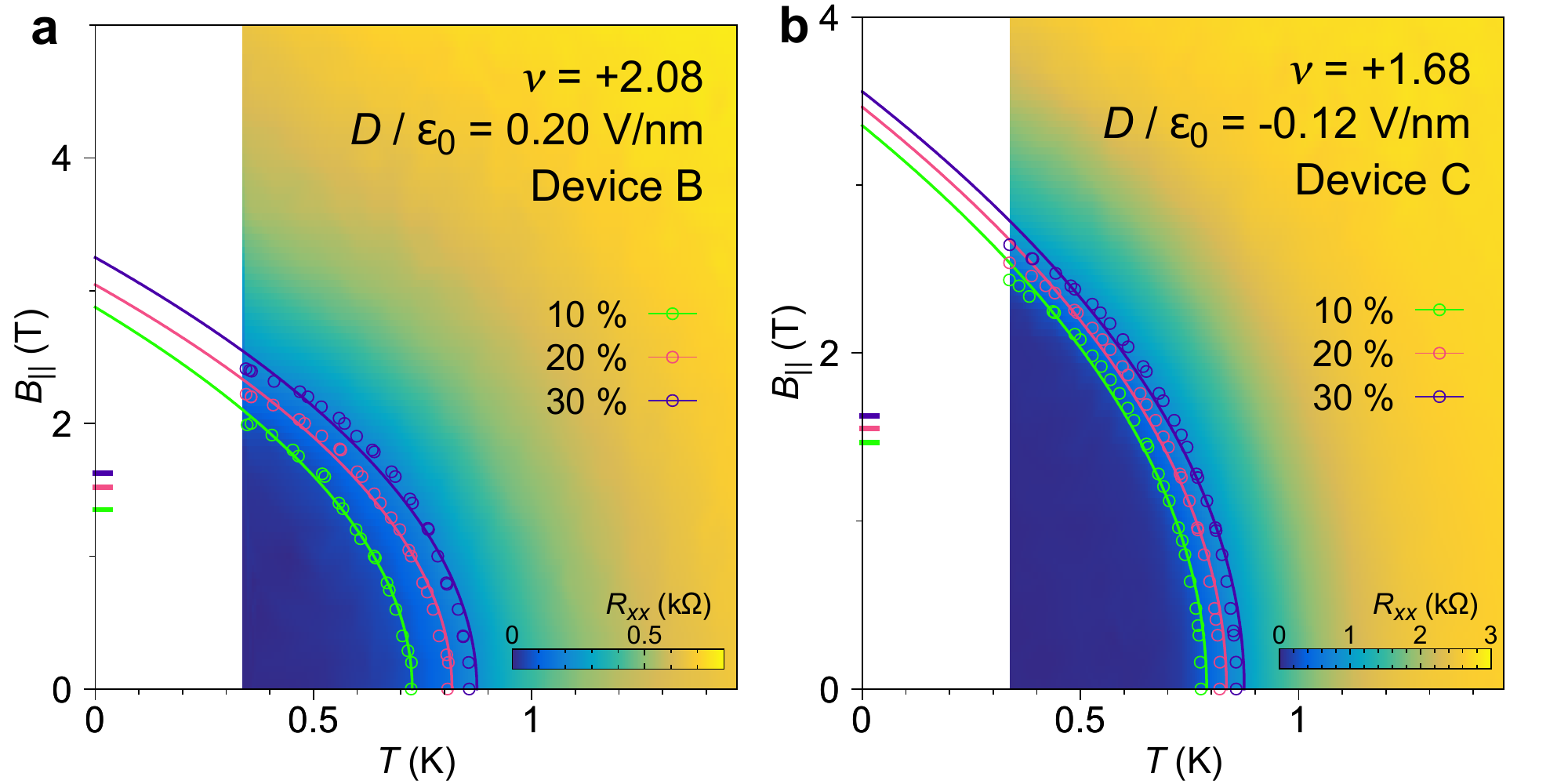}
\caption{Pauli limit violation in other devices. (a) $B_\parallel$-$T$ phase diagram of device B with twist angle $\theta\approx\SI{1.44}{\degree}$. The extracted Pauli limit violation ratios using \SI{10}{\percent}, \SI{20}{\percent} and \SI{30}{\percent} of the normal state resistance as the threshold are 2.13, 2.00, and 2.00, respectively. (b) $B_\parallel$-$T$ phase diagram of device C with twist angle $\theta\approx\SI{1.4}{\degree}$. The extracted Pauli limit violation ratios using \SI{10}{\percent}, \SI{20}{\percent} and \SI{30}{\percent} of the normal state resistance as the threshold are 2.29, 2.23, and 2.19, respectively. The solid lines show the fit to the Ginzburg-Landau expression $T\propto 1-\alpha B_\parallel^2$, and the color tickmarks at $T=0$ show the corresponding Pauli limit, the same as in Fig. 2 in the main text.}
\end{figure}

\clearpage
\begin{figure}[!ht]
\includegraphics[width=\textwidth]{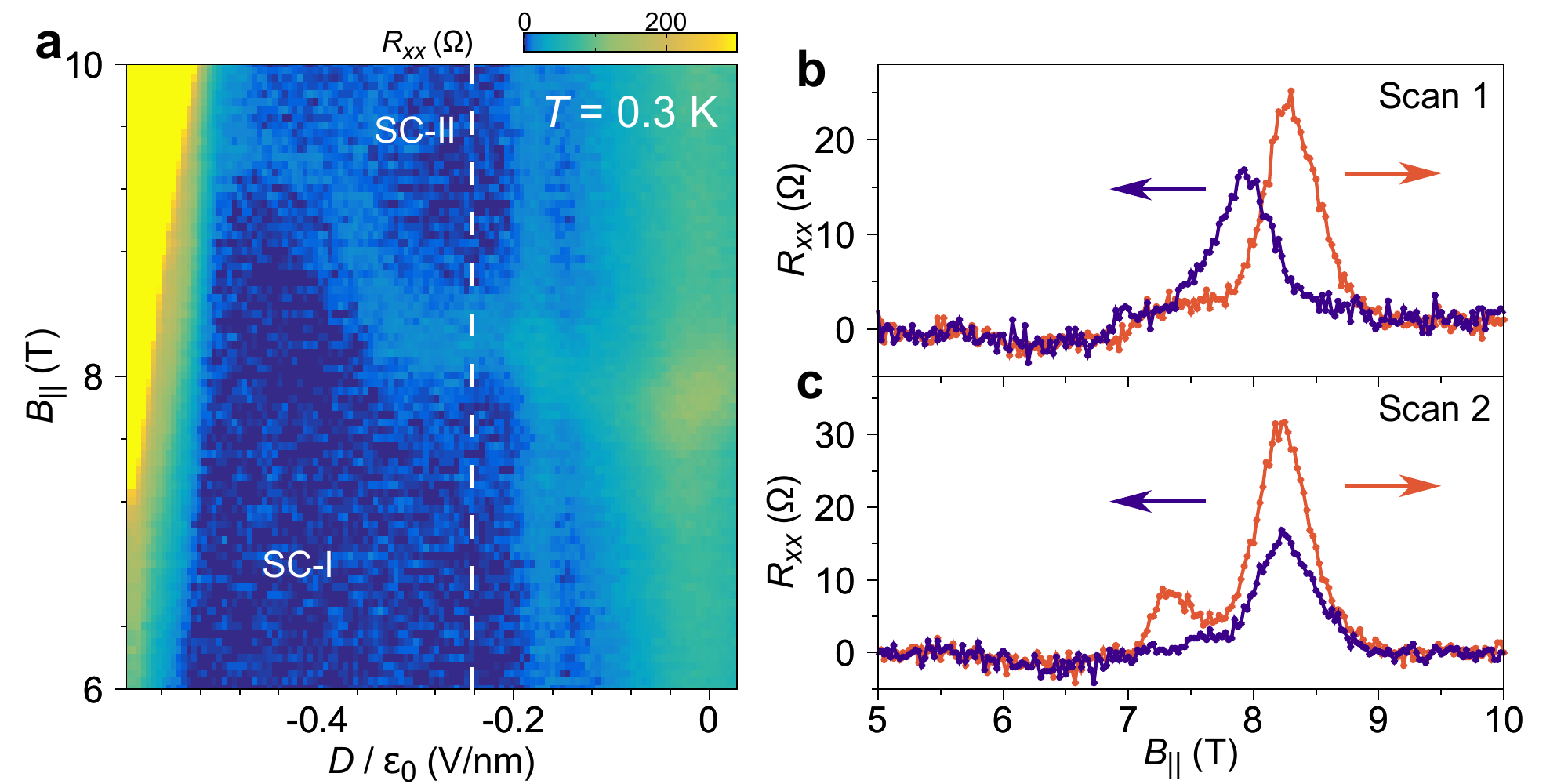}
\caption{Additional data on the high-field phases. (a) $D$-$B_\parallel$ map of resistance at a lower temperature $T=\SI{0.3}{\kelvin}$ (see Fig. 4a for comparison). The filament-like transition between SC-I and SC-II is much less pronounced. (b-c) Bidirectional sweeps in $B_\parallel$ at fixed $D$ indicated by the white dashed line in (a). The only change in measurement conditions between the two scans is a different arrangement of the BNC cables connecting to the lock-in amplifiers. Both scans are performed at \SI{0.3}{\kelvin}. }
\end{figure}


\begin{thebibliography}{10}
\expandafter\ifx\csname url\endcsname\relax
  \def\url#1{\texttt{#1}}\fi
\expandafter\ifx\csname urlprefix\endcsname\relax\def\urlprefix{URL }\fi
\providecommand{\bibinfo}[2]{#2}
\providecommand{\eprint}[2][]{\url{#2}}

\bibitem{cao_unconventional_2018}
\bibinfo{author}{Cao, Y.} \emph{et~al.}
\newblock \bibinfo{title}{Unconventional superconductivity in magic-angle
  graphene superlattices}.
\newblock \emph{\bibinfo{journal}{Nature}} \textbf{\bibinfo{volume}{556}},
  \bibinfo{pages}{43--50} (\bibinfo{year}{2018}).

\bibitem{yankowitz_tuning_2019}
\bibinfo{author}{Yankowitz, M.} \emph{et~al.}
\newblock \bibinfo{title}{Tuning superconductivity in twisted bilayer
  graphene}.
\newblock \emph{\bibinfo{journal}{Science}} \textbf{\bibinfo{volume}{363}},
  \bibinfo{pages}{1059--1064} (\bibinfo{year}{2019}).

\bibitem{lu_superconductors_2019}
\bibinfo{author}{Lu, X.} \emph{et~al.}
\newblock \bibinfo{title}{Superconductors, orbital magnets and correlated
  states in magic-angle bilayer graphene}.
\newblock \emph{\bibinfo{journal}{Nature}} \textbf{\bibinfo{volume}{574}},
  \bibinfo{pages}{653--657} (\bibinfo{year}{2019}).

\bibitem{park_tunable_2021}
\bibinfo{author}{Park, J.~M.}, \bibinfo{author}{Cao, Y.},
  \bibinfo{author}{Watanabe, K.}, \bibinfo{author}{Taniguchi, T.} \&
  \bibinfo{author}{Jarillo-Herrero, P.}
\newblock \bibinfo{title}{Tunable strongly coupled superconductivity in
  magic-angle twisted trilayer graphene}.
\newblock \emph{\bibinfo{journal}{Nature}} \textbf{\bibinfo{volume}{590}},
  \bibinfo{pages}{249--255} (\bibinfo{year}{2021}).

\bibitem{hao_electric_2021}
\bibinfo{author}{Hao, Z.} \emph{et~al.}
\newblock \bibinfo{title}{Electric field–tunable superconductivity in
  alternating-twist magic-angle trilayer graphene}.
\newblock \emph{\bibinfo{journal}{Science}} \textbf{\bibinfo{volume}{371}},
  \bibinfo{pages}{1133--1138} (\bibinfo{year}{2021}).

\bibitem{cao_correlated_2018}
\bibinfo{author}{Cao, Y.} \emph{et~al.}
\newblock \bibinfo{title}{Correlated insulator behaviour at half-filling in
  magic-angle graphene superlattices}.
\newblock \emph{\bibinfo{journal}{Nature}} \textbf{\bibinfo{volume}{556}},
  \bibinfo{pages}{80--84} (\bibinfo{year}{2018}).

\bibitem{sharpe_emergent_2019}
\bibinfo{author}{Sharpe, A.~L.} \emph{et~al.}
\newblock \bibinfo{title}{Emergent ferromagnetism near three-quarters filling
  in twisted bilayer graphene}.
\newblock \emph{\bibinfo{journal}{Science}} \textbf{\bibinfo{volume}{365}},
  \bibinfo{pages}{605--608} (\bibinfo{year}{2019}).

\bibitem{serlin_intrinsic_2020}
\bibinfo{author}{Serlin, M.} \emph{et~al.}
\newblock \bibinfo{title}{Intrinsic quantized anomalous {Hall} effect in a
  moiré heterostructure}.
\newblock \emph{\bibinfo{journal}{Science}} \textbf{\bibinfo{volume}{367}},
  \bibinfo{pages}{900--903} (\bibinfo{year}{2020}).

\bibitem{chen_evidence_2019}
\bibinfo{author}{Chen, G.} \emph{et~al.}
\newblock \bibinfo{title}{Evidence of a gate-tunable {Mott} insulator in a
  trilayer graphene moiré superlattice}.
\newblock \emph{\bibinfo{journal}{Nature Physics}}
  \textbf{\bibinfo{volume}{15}}, \bibinfo{pages}{237} (\bibinfo{year}{2019}).

\bibitem{burg_correlated_2019}
\bibinfo{author}{Burg, G.~W.} \emph{et~al.}
\newblock \bibinfo{title}{Correlated {Insulating} {States} in {Twisted}
  {Double} {Bilayer} {Graphene}}.
\newblock \emph{\bibinfo{journal}{Physical Review Letters}}
  \textbf{\bibinfo{volume}{123}}, \bibinfo{pages}{197702}
  (\bibinfo{year}{2019}).

\bibitem{shen_correlated_2020}
\bibinfo{author}{Shen, C.} \emph{et~al.}
\newblock \bibinfo{title}{Correlated states in twisted double bilayer
  graphene}.
\newblock \emph{\bibinfo{journal}{Nature Physics}}
  \textbf{\bibinfo{volume}{16}}, \bibinfo{pages}{520--525}
  (\bibinfo{year}{2020}).

\bibitem{cao_tunable_2020}
\bibinfo{author}{Cao, Y.} \emph{et~al.}
\newblock \bibinfo{title}{Tunable correlated states and spin-polarized phases
  in twisted bilayer–bilayer graphene}.
\newblock \emph{\bibinfo{journal}{Nature}} \textbf{\bibinfo{volume}{583}},
  \bibinfo{pages}{215--220} (\bibinfo{year}{2020}).

\bibitem{liu_tunable_2020}
\bibinfo{author}{Liu, X.} \emph{et~al.}
\newblock \bibinfo{title}{Tunable spin-polarized correlated states in twisted
  double bilayer graphene}.
\newblock \emph{\bibinfo{journal}{Nature}} \textbf{\bibinfo{volume}{583}},
  \bibinfo{pages}{221--225} (\bibinfo{year}{2020}).

\bibitem{regan_mott_2020}
\bibinfo{author}{Regan, E.~C.} \emph{et~al.}
\newblock \bibinfo{title}{Mott and generalized {Wigner} crystal states in {WSe}
  2 /{WS} 2 moiré superlattices}.
\newblock \emph{\bibinfo{journal}{Nature}} \textbf{\bibinfo{volume}{579}},
  \bibinfo{pages}{359--363} (\bibinfo{year}{2020}).

\bibitem{tang_simulation_2020}
\bibinfo{author}{Tang, Y.} \emph{et~al.}
\newblock \bibinfo{title}{Simulation of {Hubbard} model physics in {WSe} 2
  /{WS} 2 moiré superlattices}.
\newblock \emph{\bibinfo{journal}{Nature}} \textbf{\bibinfo{volume}{579}},
  \bibinfo{pages}{353--358} (\bibinfo{year}{2020}).

\bibitem{wang_correlated_2020}
\bibinfo{author}{Wang, L.} \emph{et~al.}
\newblock \bibinfo{title}{Correlated electronic phases in twisted bilayer
  transition metal dichalcogenides}.
\newblock \emph{\bibinfo{journal}{Nature Materials}}
  \textbf{\bibinfo{volume}{19}}, \bibinfo{pages}{861--866}
  (\bibinfo{year}{2020}).

\bibitem{lee_doping_2006}
\bibinfo{author}{Lee, P.~A.}, \bibinfo{author}{Nagaosa, N.} \&
  \bibinfo{author}{Wen, X.-G.}
\newblock \bibinfo{title}{Doping a {Mott} insulator: {Physics} of
  high-temperature superconductivity}.
\newblock \emph{\bibinfo{journal}{Reviews of Modern Physics}}
  \textbf{\bibinfo{volume}{78}}, \bibinfo{pages}{17--85}
  (\bibinfo{year}{2006}).

\bibitem{strand_transition_2010}
\bibinfo{author}{Strand, J.~D.} \emph{et~al.}
\newblock \bibinfo{title}{The {Transition} {Between} {Real} and {Complex}
  {Superconducting} {Order} {Parameter} {Phases} in {UPt3}}.
\newblock \emph{\bibinfo{journal}{Science}} \textbf{\bibinfo{volume}{328}},
  \bibinfo{pages}{1368--1369} (\bibinfo{year}{2010}).

\bibitem{schemm_observation_2014}
\bibinfo{author}{Schemm, E.~R.}, \bibinfo{author}{Gannon, W.~J.},
  \bibinfo{author}{Wishne, C.~M.}, \bibinfo{author}{Halperin, W.~P.} \&
  \bibinfo{author}{Kapitulnik, A.}
\newblock \bibinfo{title}{Observation of broken time-reversal symmetry in the
  heavy-fermion superconductor {UPt3}}.
\newblock \emph{\bibinfo{journal}{Science}} \textbf{\bibinfo{volume}{345}},
  \bibinfo{pages}{190--193} (\bibinfo{year}{2014}).

\bibitem{ran_nearly_2019}
\bibinfo{author}{Ran, S.} \emph{et~al.}
\newblock \bibinfo{title}{Nearly ferromagnetic spin-triplet superconductivity}.
\newblock \emph{\bibinfo{journal}{Science}} \textbf{\bibinfo{volume}{365}},
  \bibinfo{pages}{684--687} (\bibinfo{year}{2019}).

\bibitem{leggett_theoretical_1975}
\bibinfo{author}{Leggett, A.~J.}
\newblock \bibinfo{title}{A theoretical description of the new phases of liquid
  \${\textasciicircum}\{3\}{\textbackslash}mathrm\{{He}\}\$}.
\newblock \emph{\bibinfo{journal}{Reviews of Modern Physics}}
  \textbf{\bibinfo{volume}{47}}, \bibinfo{pages}{331--414}
  (\bibinfo{year}{1975}).

\bibitem{kitaev_unpaired_2001}
\bibinfo{author}{Kitaev, A.~Y.}
\newblock \bibinfo{title}{Unpaired {Majorana} fermions in quantum wires}.
\newblock \emph{\bibinfo{journal}{Physics-Uspekhi}}
  \textbf{\bibinfo{volume}{44}}, \bibinfo{pages}{131--136}
  (\bibinfo{year}{2001}).

\bibitem{khalaf_magic_2019}
\bibinfo{author}{Khalaf, E.}, \bibinfo{author}{Kruchkov, A.~J.},
  \bibinfo{author}{Tarnopolsky, G.} \& \bibinfo{author}{Vishwanath, A.}
\newblock \bibinfo{title}{Magic angle hierarchy in twisted graphene
  multilayers}.
\newblock \emph{\bibinfo{journal}{Physical Review B}}
  \textbf{\bibinfo{volume}{100}}, \bibinfo{pages}{085109}
  (\bibinfo{year}{2019}).

\bibitem{carr_ultraheavy_2020}
\bibinfo{author}{Carr, S.} \emph{et~al.}
\newblock \bibinfo{title}{Ultraheavy and {Ultrarelativistic} {Dirac}
  {Quasiparticles} in {Sandwiched} {Graphenes}}.
\newblock \emph{\bibinfo{journal}{Nano Letters}} \textbf{\bibinfo{volume}{20}},
  \bibinfo{pages}{3030--3038} (\bibinfo{year}{2020}).

\bibitem{lei_mirror_2020}
\bibinfo{author}{Lei, C.}, \bibinfo{author}{Linhart, L.}, \bibinfo{author}{Qin,
  W.}, \bibinfo{author}{Libisch, F.} \& \bibinfo{author}{MacDonald, A.~H.}
\newblock \bibinfo{title}{Mirror {Symmetry} {Breaking} and {Stacking}-{Shift}
  {Dependence} in {Twisted} {Trilayer} {Graphene}}.
\newblock \emph{\bibinfo{journal}{arXiv:2010.05787 [cond-mat]}}
  (\bibinfo{year}{2020}).

\bibitem{calugaru_tstg_2021}
\bibinfo{author}{Călugăru, D.} \emph{et~al.}
\newblock \bibinfo{title}{{TSTG} {I}: {Single}-{Particle} and {Many}-{Body}
  {Hamiltonians} and {Hidden} {Non}-local {Symmetries} of {Trilayer}
  {Moir}{\textbackslash}'e {Systems} with and without {Displacement} {Field}}.
\newblock \emph{\bibinfo{journal}{arXiv:2102.06201 [cond-mat]}}
  (\bibinfo{year}{2021}).

\bibitem{tinkham_introduction_2004}
\bibinfo{author}{Tinkham, M.}
\newblock \emph{\bibinfo{title}{Introduction to {Superconductivity}}}
  (\bibinfo{publisher}{Dover Publications}, \bibinfo{year}{2004}),
  \bibinfo{edition}{2nd} edn.

\bibitem{kwan_twisted_2020}
\bibinfo{author}{Kwan, Y.~H.}, \bibinfo{author}{Parameswaran, S.~A.} \&
  \bibinfo{author}{Sondhi, S.~L.}
\newblock \bibinfo{title}{Twisted bilayer graphene in a parallel magnetic
  field}.
\newblock \emph{\bibinfo{journal}{Physical Review B}}
  \textbf{\bibinfo{volume}{101}}, \bibinfo{pages}{205116}
  (\bibinfo{year}{2020}).

\bibitem{cao_nematicity_2020}
\bibinfo{author}{Cao, Y.} \emph{et~al.}
\newblock \bibinfo{title}{Nematicity and {Competing} {Orders} in
  {Superconducting} {Magic}-{Angle} {Graphene}}.
\newblock \emph{\bibinfo{journal}{arXiv:2004.04148 [cond-mat]}}
  (\bibinfo{year}{2020}).

\bibitem{chandrasekhar_note_1962}
\bibinfo{author}{Chandrasekhar, B.~S.}
\newblock \bibinfo{title}{A note on the maximum critical field of high‐field
  superconductors}.
\newblock \emph{\bibinfo{journal}{Applied Physics Letters}}
  \textbf{\bibinfo{volume}{1}}, \bibinfo{pages}{7--8} (\bibinfo{year}{1962}).

\bibitem{clogston_upper_1962}
\bibinfo{author}{Clogston, A.~M.}
\newblock \bibinfo{title}{Upper {Limit} for the {Critical} {Field} in {Hard}
  {Superconductors}}.
\newblock \emph{\bibinfo{journal}{Physical Review Letters}}
  \textbf{\bibinfo{volume}{9}}, \bibinfo{pages}{266--267}
  (\bibinfo{year}{1962}).

\bibitem{fulde_superconductivity_1964}
\bibinfo{author}{Fulde, P.} \& \bibinfo{author}{Ferrell, R.~A.}
\newblock \bibinfo{title}{Superconductivity in a {Strong} {Spin}-{Exchange}
  {Field}}.
\newblock \emph{\bibinfo{journal}{Physical Review}}
  \textbf{\bibinfo{volume}{135}}, \bibinfo{pages}{A550--A563}
  (\bibinfo{year}{1964}).

\bibitem{larkin_nonuniform_1965}
\bibinfo{author}{Larkin, A.~I.} \& \bibinfo{author}{Ovchinnikov, Y.~N.}
\newblock \bibinfo{title}{Nonuniform {State} of {Superconductors}}.
\newblock \emph{\bibinfo{journal}{Soviet Physics-JETP}}
  \textbf{\bibinfo{volume}{20}}, \bibinfo{pages}{762--770}
  (\bibinfo{year}{1965}).

\bibitem{burkhardt_fulde-ferrell-larkin-ovchinnikov_1994}
\bibinfo{author}{Burkhardt, H.} \& \bibinfo{author}{Rainer, D.}
\newblock \bibinfo{title}{Fulde-{Ferrell}-{Larkin}-{Ovchinnikov} state in
  layered superconductors}.
\newblock \emph{\bibinfo{journal}{Annalen der Physik}}
  \textbf{\bibinfo{volume}{506}}, \bibinfo{pages}{181--194}
  (\bibinfo{year}{1994}).

\bibitem{lu_evidence_2015}
\bibinfo{author}{Lu, J.~M.} \emph{et~al.}
\newblock \bibinfo{title}{Evidence for two-dimensional {Ising}
  superconductivity in gated {MoS2}}.
\newblock \emph{\bibinfo{journal}{Science}} \textbf{\bibinfo{volume}{350}},
  \bibinfo{pages}{1353--1357} (\bibinfo{year}{2015}).

\bibitem{saito_superconductivity_2016}
\bibinfo{author}{Saito, Y.} \emph{et~al.}
\newblock \bibinfo{title}{Superconductivity protected by spin–valley locking
  in ion-gated {MoS} 2}.
\newblock \emph{\bibinfo{journal}{Nature Physics}}
  \textbf{\bibinfo{volume}{12}}, \bibinfo{pages}{144--149}
  (\bibinfo{year}{2016}).

\bibitem{xi_ising_2016}
\bibinfo{author}{Xi, X.} \emph{et~al.}
\newblock \bibinfo{title}{Ising pairing in superconducting {NbSe} 2 atomic
  layers}.
\newblock \emph{\bibinfo{journal}{Nature Physics}}
  \textbf{\bibinfo{volume}{12}}, \bibinfo{pages}{139--143}
  (\bibinfo{year}{2016}).

\bibitem{avsar_colloquium_2020}
\bibinfo{author}{Avsar, A.} \emph{et~al.}
\newblock \bibinfo{title}{Colloquium: {Spintronics} in graphene and other
  two-dimensional materials}.
\newblock \emph{\bibinfo{journal}{Reviews of Modern Physics}}
  \textbf{\bibinfo{volume}{92}}, \bibinfo{pages}{021003}
  (\bibinfo{year}{2020}).

\bibitem{sigrist_introduction_2005}
\bibinfo{author}{Sigrist, M.}
\newblock \bibinfo{title}{Introduction to {Unconventional}
  {Superconductivity}}.
\newblock \emph{\bibinfo{journal}{AIP Conference Proceedings}}
  \textbf{\bibinfo{volume}{789}}, \bibinfo{pages}{165--243}
  (\bibinfo{year}{2005}).

\bibitem{khalaf_symmetry_2020}
\bibinfo{author}{Khalaf, E.}, \bibinfo{author}{Ledwith, P.} \&
  \bibinfo{author}{Vishwanath, A.}
\newblock \bibinfo{title}{Symmetry constraints on superconductivity in twisted
  bilayer graphene: {Fractional} vortices, \$4e\$ condensates or non-unitary
  pairing}.
\newblock \emph{\bibinfo{journal}{arXiv:2012.05915 [cond-mat]}}
  (\bibinfo{year}{2020}).

\bibitem{uji_magnetic-field-induced_2001}
\bibinfo{author}{Uji, S.} \emph{et~al.}
\newblock \bibinfo{title}{Magnetic-field-induced superconductivity in a
  two-dimensional organic conductor}.
\newblock \emph{\bibinfo{journal}{Nature}} \textbf{\bibinfo{volume}{410}},
  \bibinfo{pages}{908--910} (\bibinfo{year}{2001}).

\bibitem{balicas_superconductivity_2001}
\bibinfo{author}{Balicas, L.} \emph{et~al.}
\newblock \bibinfo{title}{Superconductivity in an {Organic} {Insulator} at
  {Very} {High} {Magnetic} {Fields}}.
\newblock \emph{\bibinfo{journal}{Physical Review Letters}}
  \textbf{\bibinfo{volume}{87}}, \bibinfo{pages}{067002}
  (\bibinfo{year}{2001}).

\bibitem{aoki_review_2019}
\bibinfo{author}{Aoki, D.}, \bibinfo{author}{Ishida, K.} \&
  \bibinfo{author}{Flouquet, J.}
\newblock \bibinfo{title}{Review of {U}-based {Ferromagnetic}
  {Superconductors}: {Comparison} between {UGe2}, {URhGe}, and {UCoGe}}.
\newblock \emph{\bibinfo{journal}{Journal of the Physical Society of Japan}}
  \textbf{\bibinfo{volume}{88}}, \bibinfo{pages}{022001}
  (\bibinfo{year}{2019}).

\bibitem{christos_superconductivity_2020}
\bibinfo{author}{Christos, M.}, \bibinfo{author}{Sachdev, S.} \&
  \bibinfo{author}{Scheurer, M.~S.}
\newblock \bibinfo{title}{Superconductivity, correlated insulators, and
  {Wess}–{Zumino}–{Witten} terms in twisted bilayer graphene}.
\newblock \emph{\bibinfo{journal}{Proceedings of the National Academy of
  Sciences}} \textbf{\bibinfo{volume}{117}}, \bibinfo{pages}{29543--29554}
  (\bibinfo{year}{2020}).

\bibitem{cea_coulomb_2021}
\bibinfo{author}{Cea, T.} \& \bibinfo{author}{Guinea, F.}
\newblock \bibinfo{title}{Coulomb interaction, phonons, and superconductivity
  in twisted bilayer graphene}.
\newblock \emph{\bibinfo{journal}{arXiv:2103.01815 [cond-mat]}}
  (\bibinfo{year}{2021}).

\bibitem{park_flavour_2020}
\bibinfo{author}{Park, J.~M.}, \bibinfo{author}{Cao, Y.},
  \bibinfo{author}{Watanabe, K.}, \bibinfo{author}{Taniguchi, T.} \&
  \bibinfo{author}{Jarillo-Herrero, P.}
\newblock \bibinfo{title}{Flavour {Hund}'s {Coupling}, {Correlated} {Chern}
  {Gaps}, and {Diffusivity} in {Moiré} {Flat} {Bands}}.
\newblock \emph{\bibinfo{journal}{arXiv:2008.12296 [cond-mat]}}
  (\bibinfo{year}{2020}).

\end{thebibliography}
\end{document}